\documentclass[iopart,groupedaddress,superscriptaddress,nofootinbib,includegraphics]{revtex4-1}

\usepackage[pdftex,linkcolor=blue,citecolor=blue,urlcolor=blue,colorlinks]{hyperref}
\usepackage[dvipsnames]{xcolor}
\usepackage{graphicx,epsfig}
\usepackage{color}
\usepackage{cancel}
\usepackage{dsfont}
\usepackage{inputenc}
\usepackage{amsmath}
\usepackage{amsfonts}
\usepackage{fancyhdr}
\usepackage{ulem}

\newcommand{\ket}[1]{\ensuremath{\left\vert #1 \right\rangle}}
\newcommand{\bra}[1]{\ensuremath{\left\langle #1 \right\vert}}

\newcommand{\fref}[1]{Fig.~\ref{#1}}

\newcommand{\cref}[1]{chapter~\ref{#1}}
\newcommand{\Cref}[1]{Chapter~\ref{#1}}

\begin{document}
\title{High-fidelity multiqubit Rydberg gates via two-photon adiabatic rapid passage}
\author{G. Pelegr\'i}
\affiliation{Department of Physics and SUPA, University of Strathclyde, Glasgow G4 0NG, UK}
\author{A. J. Daley}
\affiliation{Department of Physics and SUPA, University of Strathclyde, Glasgow G4 0NG, UK}
\author{J. D. Pritchard}
\affiliation{Department of Physics and SUPA, University of Strathclyde, Glasgow G4 0NG, UK}

%\email{g.pelegri@strath.ac.uk} %optional

\begin{abstract}
We present a robust protocol for implementing high-fidelity multiqubit controlled phase gates $(C^kZ)$ on neutral atom qubits coupled to highly excited Rydberg states. Our approach is based on extending adiabatic rapid passage to two-photon excitation via a short-lived intermediate excited state common to alkali-atom Rydberg experiments, accounting for the full impact of spontaneous decay and differential AC Stark shifts from the complete manifold of hyperfine excited states. We evaluate and optimise gate performance, concluding that for Cs and currently available laser frequencies and powers, a $CCZ$ gate with fidelity $\mathcal{F}>0.995$ for three qubits and $CCCZ$ with $\mathcal{F}>0.99$ for four qubits is attainable in $\sim 1.8$~$\mu$s via this protocol. Higher fidelities are accessible with future technologies, and our results highlight the utility of neutral atom arrays for the native implementation of multiqubit unitaries.
\end{abstract}
\pacs{}
\maketitle

\section{Introduction}
Over the last decade, trapped neutral atoms have emerged as one of the most promising platforms for quantum information processing \cite{saffman2010quantum,adams2019rydberg,henriet2020quantum,morgado2021quantum}.  Several experiments have demonstrated reliable single atom loading, initialisation and readout in configurable 1D \cite{endres2016atom}, 2D \cite{barredo2016atom, deMello2019defect} and 3D \cite{kumar2018sorting,barredo2018synthetic} optical arrays formed by up to a few hundred sites \cite{scholl2020programmable,bluvstein2021controlling}, without fundamental limitations to further scaling.  Control over high-fidelity single-qubit gates for neutral atoms, with qubits encoded in long-lived hyperfine states, has been demonstrated in several settings \cite{xia2015randomized,wang2016single,sheng2018high,wu2019stern}. By exciting the atoms to strongly interacting Rydberg states, it is then possible to implement controlled-NOT or controlled-phase two-qubit gates \cite{lukin2001dipole,jaksch2000fast}, which are key for the realisation of a universal gate set for digital quantum computing \cite{nielsen2002quantum}. Although the coherent excitation of the ground-Rydberg transitions poses several challenges \cite{deLeseluc2018analysis} that had limited the fidelity of two-qubit gates for a number of years \cite{wilk2010entanglement,maller2015rydberg,jau2016entangling,zeng2017entangling,picken2018entanglement}, recent experiments have overcome many of these difficulties through suppression of technical noise \cite{levine2018high} and incorporation of optimal control techniques \cite{Goerz14,Levine2019parallel,saffman2020symmetric,pagano2022error,jandura2022time}, reporting on the preparation of Bell states with fidelities $\mathcal{F}\simeq 0.97$ using alkali atoms in 1D arrays \cite{Levine2019parallel} and $\mathcal{F}\simeq 0.89$ in 2D arrays \cite{graham2019rydberg}, and $\mathcal{F}>0.99$ with alkaline-earth atoms \cite{madjarov2020high}.

%Following our conversation I propose we aim to use this paragraph to say multi-qubit gates are great, but challenging to realise without introducing complex requirements such as local addressing, engineering of the interactions, or limiting geometries. Our gate then looks "simple"
Beyond the long-lived encoding in physically identical qubits, a crucial potential advantage of the neutral atom platform is the ability to implement native multiqubit operations %, exploiting the strong, controllable long-range interactions   
\cite{brion07a,isenhower11,beterov18,shi18,su18,li21,young21,wu10a,su18a,khazali20,rasmussen20,muller2009mesoscopic,molmer11,petrosyan16} or %. for efficient digital quantum computation, or digital simulation through engineering
multi-body spin interactions \cite{weimer10,weimer11}, including for quantum annealing \cite{glaetzle17} and optimisation \cite{dlaska21,lechner15}. Various approaches have made progress towards mulitply-controlled three-qubit gates, making use of local addressing \cite{brion07a,isenhower11}, geometric arrangement \cite{shi18}, use of multiple Rydberg states \cite{khazali20,wu10a}, or tuning of the dipole-dipole interaction potentials using dc electric fields \cite{beterov18} or microwave dressing \cite{young21,tanasittikosol11,sevincli14}. However, while initial demonstrations of three-qubit gates have been made \cite{Levine2019parallel}, realising robust and high-fidelity multi-qubit gates with more than a single control qubit remains a challenge due to limited Rydberg lifetimes, collective excitation effects and finite qubit interaction strengths.
%. This approach can be extended to recover effective all-to-all connectivity for quantum annealing \cite{glaetzle17} and optimisation \cite{dlaska21} using the LHZ encoding scheme \cite{lechner15}. 

%MAKE THIS STRONGER!!! Emphasis multply controlled (ref only multiple control gates) - keep "parallel"
In this work, we show that adiabatic rapid passage (ARP) techniques can be extended to provide high-fidelity multiply-controlled phase gates ($C^{k}Z$). Our protocol is tailored to the case of two-photon excitation via a rapidly decaying intermediate excited state, as is typical of experiments using alkali atoms. This scheme is robust to experimental imperfections and requires only global addressing of the qubits within the blockade regime. These multi-qubit gates have the potential to substantially optimise quantum circuit implementation \cite{saffman2010quantum,morgado2021quantum,molmer11,petrosyan16}, as well as quantum error correction \cite{auger17}. 

Our gate sequences are designed using an exact model that accounts for the effects of spontaneous decay and AC Stark shifts from the full manifold of hyperfine excited states for experimentally feasible parameters, providing a realistic evaluation of the intrinsic gate fidelities attainable for Cs atoms prepared in the motional ground state using current technologies. Starting from analytical pulse profiles for $CCZ$ on three qubits with $\mathcal{F}\sim0.993$, we apply optimal control using a variant of the dressed chopped random basis (dCRAB) method \cite{caneva2011chopped,Rach15} to design global laser pulses achieving a fidelity of $\mathcal{F}\sim0.995$, outperforming the equivalent circuit decomposition into two-qubit gates for $\mathcal{F}_{2Q}<0.999$. Extension of this method to four qubits yields a $CCCZ$ gate with $\mathcal{F}>0.99$ without optimisation, providing a viable route to demonstrating efficient and scalable digital computing on neutral atom arrays.

\section{$C^{k}Z$ gate via two-photon adiabatic rapid passage}
\begin{figure}[t!]
\centering
\includegraphics[width=17.2cm]{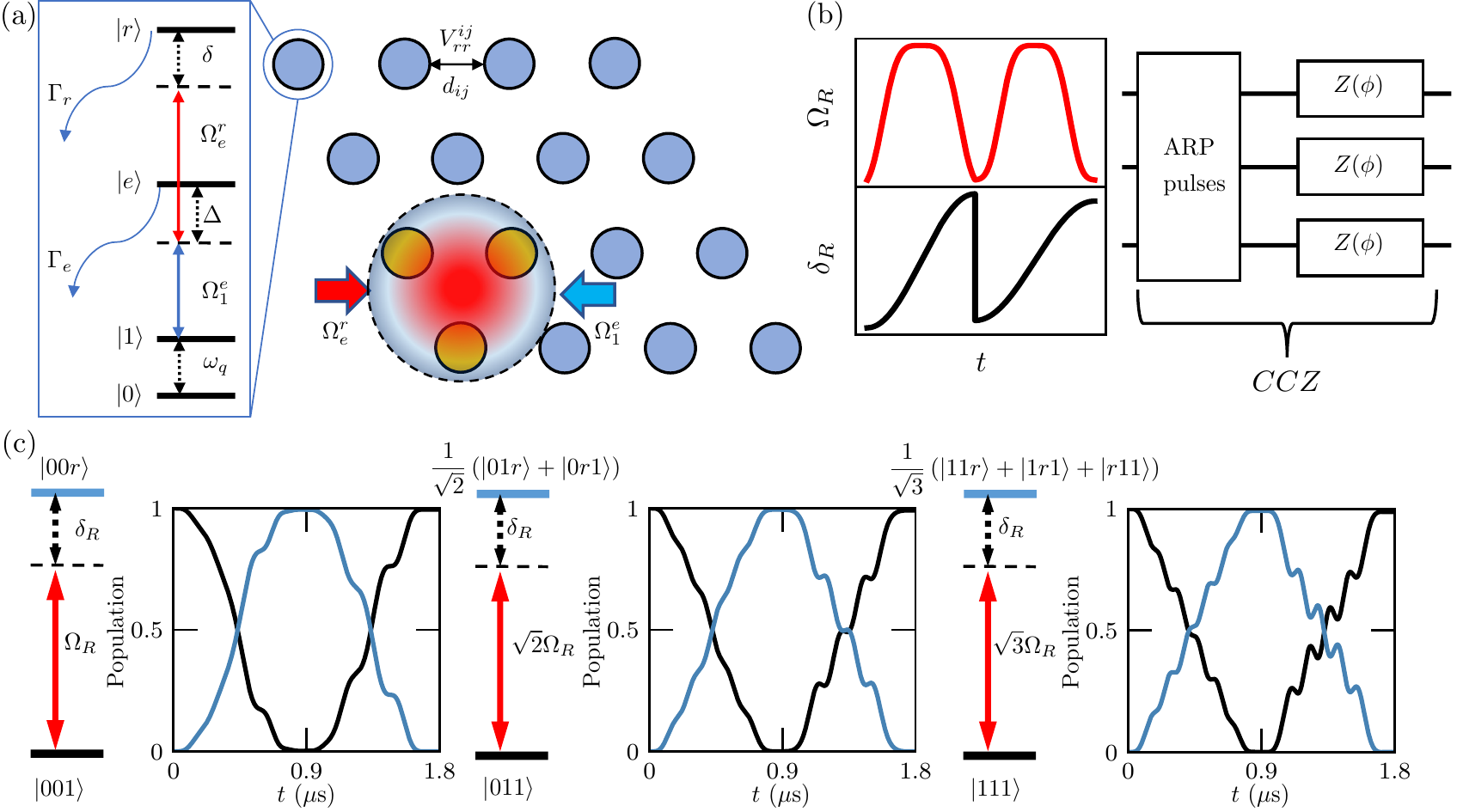}
\caption{a) Triangular lattice formed by trapped neutral atoms separated by distances $d_{ij}$ and experiencing pairwise Rydberg interactions of strength $V_{rr}^{ij}$. The inset depicts the atomic states and parameters considered in the single-atom Hamiltonian \eqref{Ham_1atom_full}. Any three nearest-neighbour atoms can be globally illuminated with laser pulses that realise the two-photon ARP $CCZ$ gate protocol by providing the drive from $\ket{1}$ to $\ket{r}$ through the intermediate state $\ket{e}$.  b) Quantum circuit representation of the pulses used to implement the $CCZ$ gate. A schematic representation of the $\Omega_R(t)$ and $\delta_R(t)$ ARP pulses is also included. c) Sketches of the effective two-level systems involved in the ARP processes for the $CCZ$ gate. For each two-level system, the time evolution of the ground (black lines) and excited (blue lines) states when the ARP pulses $\Omega_R(t),\delta_R(t)$ shown in Fig. \ref{gates_normal}(a) are applied is shown.}
\label{levelscheme}
\end{figure}
The $C^{k}Z$ gate is implemented on an ensemble of $k+1$ alkali atoms, each of which represents a qubit with the computational states $\ket{0}$ and $\ket{1}$ encoded in two hyperfine ground states with frequency separation $\omega_q$.  As illustrated in \fref{levelscheme}(a), the state $\ket{1}$ is coupled to a Rydberg state $\ket{r}$ via a two-photon transition through an intermediate excited state $\ket{e}$, which is resolved into its hyperfine components $\ket{f_e,m_{f_e}}$. The single-atom Hamiltonian describing this excitation on atom $i$ is given by
\begin{equation} \label{Ham_1atom_full}
\begin{split}
\frac{\mathcal{H}}{\hbar} &=\sum_{f_e,m_{f_e}} \left[\frac{\Omega_1^{ f_e,m_{f_e}}}{2} \Big(\ket{1}\bra{f_e,m_{f_e}} +\ket{f_e,m_{f_e}}\bra{1}\Big)
+\frac{\Omega_{f_e,m_{f_e}}^{r}}{2}\Big( \ket{f_e,m_{f_e}}\bra{r}+\ket{r}\bra{f_e,m_{f_e}}\Big)\right.\\\\
& -\Delta_{f_e,m_{f_e}}\ket {f_e,m_{f_e}}\bra{ f_e,m_{f_e}}\Bigg]-\delta\vert r \rangle \langle r \vert,
\end{split}
\end{equation}
where $\Omega_1^{ f_e,m_{f_e}}$ and $\Omega_{f_e,m_{f_e}}^{r}$ are the Rabi frequencies of the couplings from $\ket{1}\rightarrow\ket{f_e,m_{f_e}}$ and from $\ket{f_e,m_{f_e}}\rightarrow\ket{r}$,  $\Delta_{f_e,m_{f_e}}=\Delta-E(f_e,m_{f_e})$ represent the intermediate-state detunings composed by the centre of mass laser detuning $\Delta$ and the hyperfine splitting $E(f_e,m_{fe})$,  and $\delta$ is the two-photon detuning. We consider the far off-resonant limit $\vert\Delta\vert \gg \vert\Omega_1^{ f_e,m_{f_e}}\vert, \vert\Omega_{f_e,m_{f_e}}^{r}\vert$, for which the coupling from $\ket{1}$ to $\ket{r}$ can be described in terms of an effective two-level system with Rabi frequency 
\begin{equation}
\Omega_R=\displaystyle\sum_{f_e,m_{f_e}}\frac{\Omega_1^{ f_e,m_{f_e}}\Omega_{f_e,m_{f_e}}^{r}}{2\Delta_{f_e,m_{f_e}}}
\end{equation}
and detuning
\begin{equation}
\delta_R=\delta+\Delta_1-\Delta_r, 
\label{deltaRdef}
\end{equation}
where
\begin{equation}
\Delta_1=\displaystyle\sum_{f_e,m_{f_e}}\frac{\vert\Omega_1^{ f_e,m_{f_e}}\vert^2}{4\Delta_{f_e,m_{f_e}}}, \Delta_r=\displaystyle\sum_{f_e,m_{f_e}}\frac{\vert\Omega^r_{ f_e,m_{f_e}}\vert^2}{4\Delta_{f_e,m_{f_e}}} 
\end{equation}
are the AC Stark shifts of the states $\ket{1}$ and $\ket{r}$, respectively. 

To account for intrinsic gate errors arising from spontaneous decay from the excited state and Rydberg state with decay rates $\Gamma_e$ and $\Gamma_r$ respectively, we model the system evolution using the dissipative Schr\"odinger equation by introducing effective single atom terms 
\begin{equation}
\mathcal{H}' =-i\hbar\displaystyle\sum_{f_e,m_{f_e}} \left[ \frac{\Gamma_e}{2}\ket {f_e,m_{f_e}}\bra{ f_e,m_{f_e}}+\frac{\Gamma_r}{2}\vert r \rangle \langle r \vert\right].
\end{equation}
This approach provides a lower bound on the estimated fidelity as the non-Hermitian terms in the Hamiltonian cause loss of population from the computational basis, neglecting the small probability of atoms returning to the logical qubit levels. 
\\
The Rydberg states experience dipole-induced pairwise interactions described by Hamiltonain
\begin{equation}
\mathcal{H}_\mathrm{dd}=\displaystyle\sum_{j<i}\hbar V_{rr}^{ij}\vert r_ir_j\rangle\langle r_ir_j\vert, 
\end{equation}
where the strength $V_{rr}^{ij}$ depends on the separation $d_{ij}$ between the atoms $i$ and $j$ and their orientation with respect to the quantisation axis.  In the fully blockaded regime, which is fulfilled when $\vert V_{rr}^{ij}\vert\gg \vert\Omega_R\vert$ for all pairs of atoms, the ensemble can only host one Rydberg excitation. Under this condition, we seek to apply a sequence of global pulses which realises a $C^{k}Z$ gate, given by the unitary matrix
\begin{equation}
U_{C^{k}Z}=2\left(\otimes_{k+1}\ket{0} \otimes_{k+1}\bra{0}\right)-\mathds{I}.
\end{equation}
The main difficulty in finding such a pulse sequence resides in the fact that $(k+1)$-body states with different numbers $n$ of atoms in the computational state $\ket{1}$ are coupled to the manifold of singly-excited Rydberg states with different effective Rabi frequencies $\sqrt{n}\Omega_R$. To circumvent this issue we apply two consecutive ARP pulses, which can achieve full population transfer in the resulting effective two-level systems for a wide range of Rabi frequencies. The required $\pi$ phases at the end of the sequence can be imprinted by inverting the sign of the detuning at the start of the second ARP pulse \cite{beterov2020application}.
\section{$CCZ$ gate}
As a specific example of the two-photon ARP $C^{k}Z$ gate protocol, we analyse in detail the implementation of a $CCZ$ gate.  As illustrated in Fig. \ref{levelscheme}(a), the three atoms involved in the gate are arranged in a triangle with equal side lengths $d$ and experience the same pairwise Rydberg interaction strength $V_{rr}$. In a first approach, we realise the $CCZ$ gate by applying two consecutive global ARP pulses of duration $T_1$ and $T_2$, with the effective Rabi frequency and detuning varying in time according to the analytical functions \cite{saffman2020symmetric} 
\begin{align}
\Omega_R(t)&= 
      \Omega_R^{0}\left[ \frac{e^{-\left(t-T_i/2\right)^4/\tau_i^4}-e^{-\left(T_i/2\tau_i\right)^4}}{1-e^{-\left(T_i/2\tau_i\right)^4}}\right], \label{OmegaR}\\
\delta_R(t)&= - \delta_R^i\cos\left(\frac{\pi t}{T_i}\right), \label{deltaR}
\end{align}
with $i=1,2$ for the first and second pulse, respectively. In Fig. \ref{levelscheme}(c) we represent schematically the relevant states of the different two-level systems over which the ARP pulses act, and for each of them we plot the time evolution of the ground (black lines) and excited (blue lines) states for the ARP pulses given by Eqs.  \eqref{OmegaR} and \eqref{deltaR}. According to Eq. \eqref{deltaRdef}, the effective detuning $\delta_R(t)$ can be controlled through $\delta(t)$, which is the two-photon detuning resulting from the sum of the detunings of the lasers driving the $\ket{1} \leftrightarrow \ket{e}$ and $\ket{e} \leftrightarrow \ket{r}$ transitions. Therefore, this parameter can be modified in time as desired by chirping the
frequency of either laser. In order to obtain the desired shape for $\Omega_R(t)$ for a two-photon transition, we fix the intermediate state detuning $\Delta$ and the intensity $I_{e}^{r}$ of the laser driving the $\ket{e}\leftrightarrow\ket{r}$ excitation and vary the intensity of the laser coupling the states $\ket{1}$ and $\ket{e}$ as $I_{1}^{e}(t)\propto {{\Omega_R^2(t)}}$. This time-dependent laser intensity induces a differential AC Stark shift between the states $\ket{0}$ and $\ket{1}$, which after the application of the ARP pulses results in a relative phase error on each qubit. As shown in \fref{levelscheme}(b), to compensate for this effect after the ARP pulses we apply a global single-qubit rotation $Z(\phi)$, where $\phi$ is chosen to maximise the gate fidelity. In the following we assume this single-qubit rotation to be perfect as it can be implemented as an instantaneous phase-step on all future qubit pulses or using control pulses with errors below $10^{-4}$ ~\cite{sheng2018high}. 

We compute the gate fidelity using the limited tomography prescription \cite{Levine2019parallel,figgatt2017complete}
\begin{equation}
\mathcal{F}=\left\vert \langle \Psi \vert U_{CCZ}^{\dagger}U \vert \Psi\rangle \right\vert^2,
\label{fidelityformula}
\end{equation}
where $U_{CCZ}$ is the ideal gate matrix, $U$ is the non-unitary operator resulting from time evolution under the Hamiltonian $\mathcal{H}_\mathrm{tot}=\mathcal{H}+\mathcal{H}_\mathrm{dd}+\mathcal{H}'$ and the application of the global rotation $Z(\phi)$, and the initial state ${\ket{\Psi}=1/\sqrt{8}\sum_{ijk=0,1}\ket{ijk}}$ is chosen as a symmetric superposition of all possible computational basis states $\ket{ijk}$, making the fidelity sensitive to both population leakage and relative phase errors for each computational state. Note that since each of the computational states evolves in an independent Hilbert space under the application of $\mathcal{H}_\mathrm{tot}$ there are no artificial errors caused by interference between the different states forming the superposition $\ket{\Psi}$.

\begin{figure}[t!]
\includegraphics[width=17.2 cm]{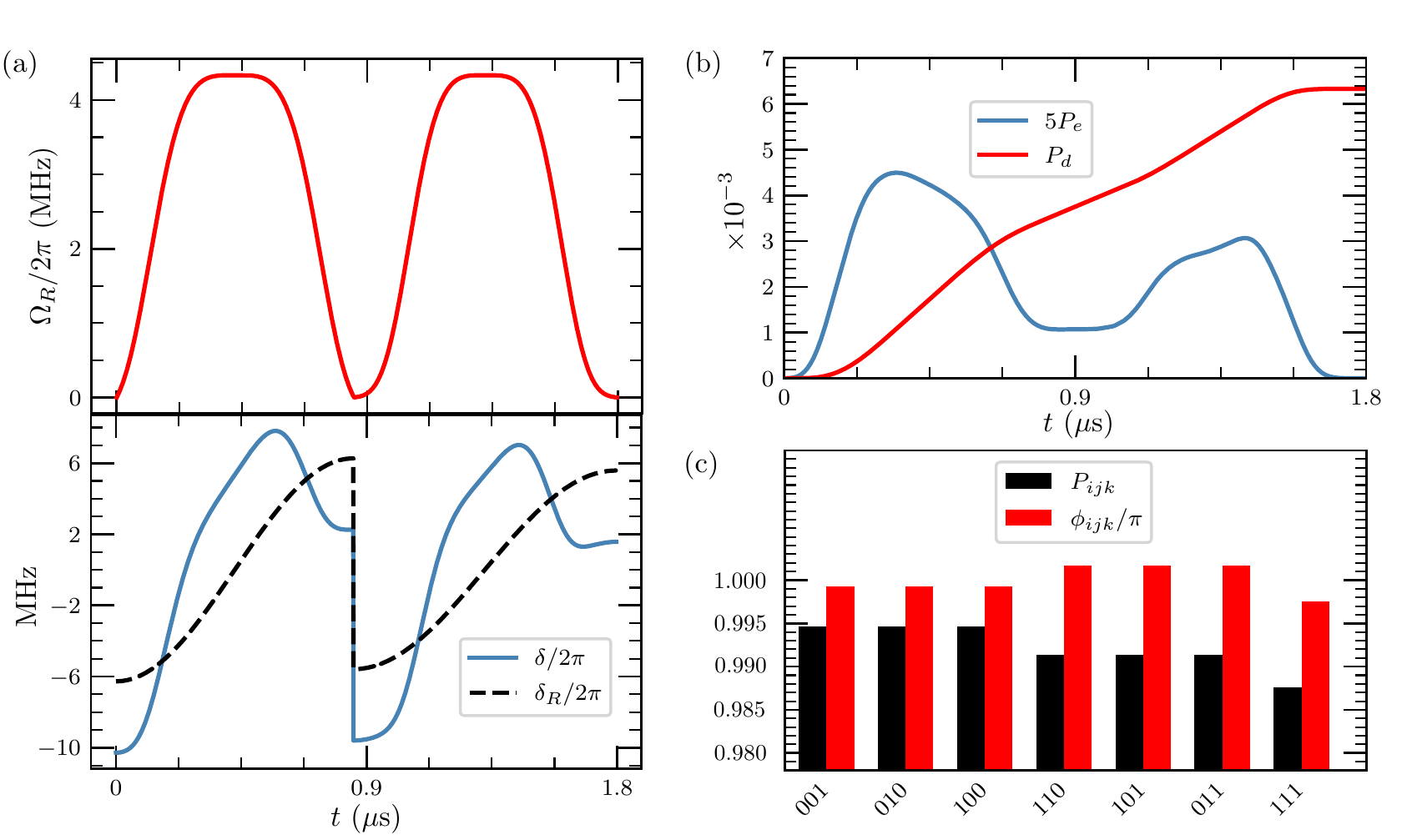}
\caption{Two-photon ARP $CCZ$ gate with optimal analytical pulses. (a) Time dependence of the Rabi frequency $\Omega_R(t)$ (upper plot) and the two-photon detuning $\delta(t)$ and the effective detuning $\delta_R(t)$ (lower plot) (b) Time evolution of the norm loss and intermediate-state population (magnified by a factor of 5) for an initial state consisting of an equal superposition of all computational states, $\ket{\Psi}=1/\sqrt{8}\sum_{ijk=0,1}\ket{ijk}$. (c) Output state population (black bars) and phase (red bars) for all the input states coupled to the Rydberg manifold after the application of the gate protocol. In all plots the parameters are $V_{rr}/2\pi=608$~MHz, $\Delta/2\pi=16.300$~GHz, $T_1=0.852$~$\mu$s,  $T_2=0.944$~$\mu$s,  $\tau_1=2.59T_1$, $\tau_2=3.19T_2$, $\delta_R^1/2\pi=6.270$~MHz, $\delta_R^2/2\pi=5.586$~MHz, and $\Omega_R^0/2\pi=4.334$, which corresponds to laser intensities $I_{1}^{e}=0.390$~mW/$\mu\text{m}^2$ and $I_{e}^{r}=6.236$~mW/$\mu\text{m}^2$.}
\label{gates_normal}
\end{figure}

We consider the explicit case of Cs atoms trapped in a triangular array separated by a distance $d=4$~$\mu$m at a temperature $T=0$ K and with a quantisation axis defined by a 10~G bias field perpendicular to the array plane.  The hyperfine ground states $\ket{0}=\ket{6S_{1/2},f=3,m_f=0}$ and $\ket{1}=\ket{6S_{1/2},f=4,m_f=0}$ embody the computational basis, and we use circularly polarised laser beams propagating parallel to the quantisation axis to couple $\ket{1}$ via the intermediate state $\ket{e}=\ket{7P_{1/2}}$ to the Rydberg state $\ket{r}=\ket{82S_{1/2},m_j=-1/2}$, for which the interaction strength is $V_{rr}/2\pi=608$~MHz.  We combine open-source packages to compute the Rydberg-Rydberg interactions \cite{weber2017calculation} and the hyperfine-resolved atom-light couplings \cite{sibalic17}.

We then optimise gate fidelity for experimentally achievable parameters, applying the realistic constraint of fixed power of $\sim 290$ mW focused to $\sim3$ $\mu$m at the center of the triangle,  corresponding to an intensity at the atom positions of $I_{e}^{r}=6.236$ mW/$\mu\text{m}^2$, and numerically optimizing intermediate state detuning, pulse parameters $T_i$, $\tau_i$ and $\delta_R^i$ ($i=1,2$),  and intensity of the laser from $\ket{1}\rightarrow\ket{e}$. To accelerate the optimisation, we use a reduced model where the dissipative contribution of the intermediate states is effectively incorporated into the Rydberg level (see Appendix \ref{reducedmodel_appendix}), and then use the optimal parameters found with this procedure to compute the time evolution with the full model.

In \fref{gates_normal} we illustrate the $CCZ$ gate sequence corresponding to the optimal parameters $\Delta/2\pi=16.3$ GHz, $T_1=0.852$~$\mu$s,  $T_2=0.944$~$\mu$s,  $\tau_1=2.59T_1$, $\tau_2=3.19T_2$, $\delta_R^1/2\pi=6.270$~MHz, $\delta_R^2/2\pi=5.586$~MHz, and $I_{1}^{e}=0.390$ mW/$\mu\text{m}^2$, for which $\Omega_R^0/2\pi=4.334$~MHz.  The fidelity obtained with the full model is $\mathcal{F}=0.9932$.  \fref{gates_normal}(a) shows the time dependence of the effective Rabi frequency $\Omega_R(t)$ in the upper plot, while the two-photon detuning $\delta(t)$ and the effective detuning $\delta_R(t)$ seen by the atoms are shown in the lower plot. In \fref{gates_normal}(b) we plot the time dependence of the population of the intermediate states $P_e$ and the leakage of population to hyperfine ground states outside of the computational basis $P_d=1-\left\langle \Psi \vert \Psi\right \rangle$ for an initial state consisting of an equal superposition of all computational states, $\ket{\Psi}=1/\sqrt{8}\sum_{ijk=0,1}\ket{ijk}$. In \fref{gates_normal}(c) we display the final population $P_{ijk}=8\vert\left\langle ijk \vert U\vert\ \Psi\right\rangle \vert^2$ and phase $\phi_{ijk}=\text{arg}\{\left\langle ijk \vert U\vert \Psi\right\rangle\}$ of all the input states other than the uncoupled state $\ket{000}$ after the application of the $CCZ$ gate. The state with the worst fidelity is $\ket{111}$, for which the output population is $P_{111}=0.988$.

\begin{figure}[t!]
\includegraphics[width=17.2cm]{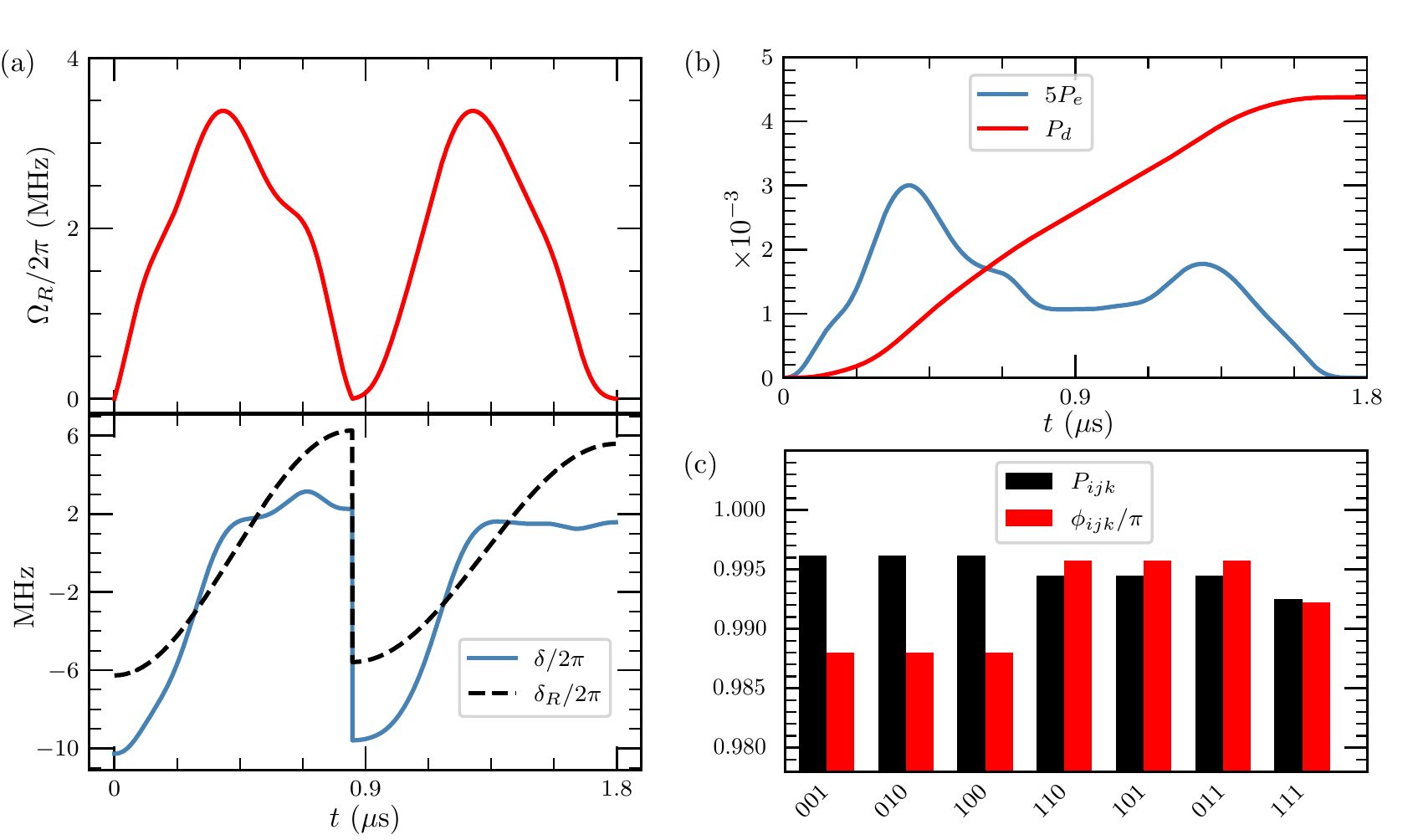}
\caption{Two-photon ARP $CCZ$ gate with dCRAB-optimised pulses. (a) Time dependence of the Rabi frequency $\Omega_R(t)$ (upper plot) and the two-photon detuning (lower plot). (b) Time evolution of the norm loss and intermediate-state population (magnified by a factor of 5) for an initial state consisting of an equal superposition of all computational states, $\ket{\Psi}=1/\sqrt{8}\sum_{ijk=0,1}\ket{ijk}$. (c) Population transfer (black bars) and phase imprinting (red bars) for all the input states coupled to the Rydberg manifold after the application of the gate protocol. In all plots the parameters are the same as in \fref{gates_normal} except for the maximum intensity of the laser driving the $\ket{1}\leftrightarrow\ket{e}$ excitation, which is reduced to $I_{1}^{e}=0.235$ mW/$\mu\text{m}^2$.}
\label{gates_optimal}
\end{figure}

\section{Pulse optimisation}
To further enhance the ARP $CCZ$ performance we apply a simplified version of the dCRAB technique \cite{caneva2011chopped,Rach15} (see Appendix \ref{optimisation} for details) to optimise the shape of the effective Rabi frequency $\Omega_R(t)$. We find a pulse shape which leads to an improvement of the fidelity to $\mathcal{F}=0.9954$. \fref{gates_optimal}(a) shows the time dependence of the dCRAB-optimised effective Rabi frequency and two-photon detuning. Besides yielding an increase in the fidelity,  this sequence reduces the peak effective Rabi frequency to $\Omega_R^\mathrm{max}/2\pi=3.38$~MHz, which corresponds to a laser intensity $I_{1}^{e}=0.235$ mW/$\mu\text{m}^2$.  In \fref{gates_optimal}(b) we plot the time evolution of $P_e$ and $P_d$ for the initial state $\ket{\Psi}=1/\sqrt{8}\sum_{ijk=0,1}\ket{ijk}$. We observe that the pulse shaping reduces the excitation of the intermediate states and the leakage to hyperfine ground states outside of the computational basis. Finally,  \fref{gates_optimal}(c) shows the output populations and phases for all of the computational states after applying the dCRAB-optimised gate. Comparing \fref{gates_optimal}(b) and \fref{gates_optimal}(c), we observe that dCRAB increases the output population for all the input states, with the worst population transfer being $P_{111} =0.993$ in this case. Although the output phase distribution for the dCRAB pulses shown in \fref{gates_optimal}(c) is further from the ideal case of a uniform $\pi$ phase across all input states than the one corresponding to the analytical pulses shown in \fref{gates_normal}(c), the gain in population transfer outweighs this effect. Note that the measure of the fidelity \eqref{fidelityformula} that we employ penalises both population loss and phase errors.
\section{$CCZ$ Gate robustness}
\begin{figure}[t!]
\includegraphics[width=17.2 cm]{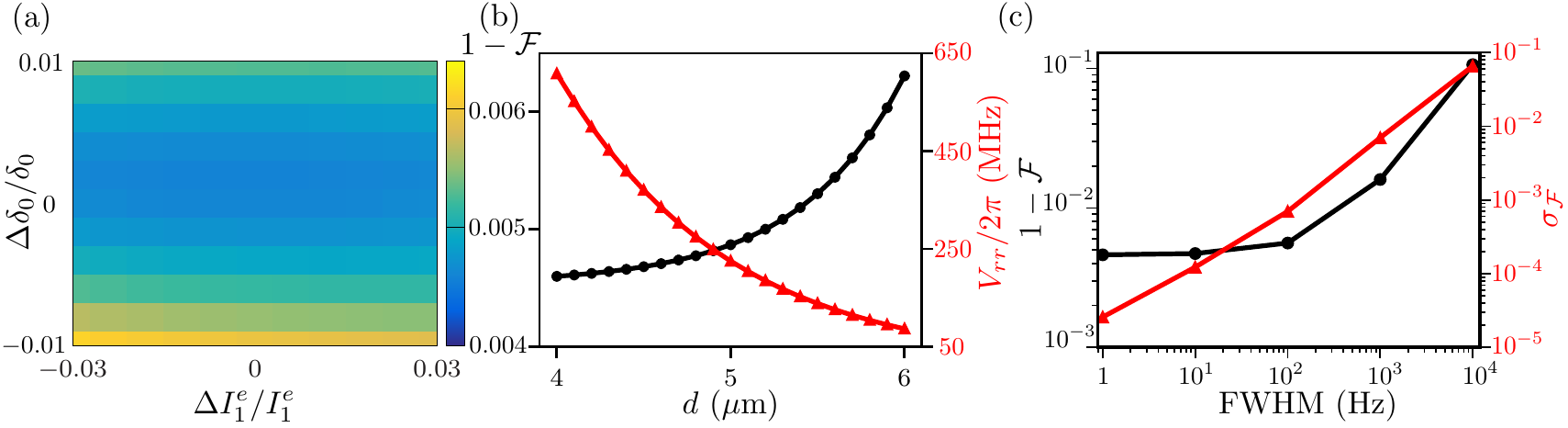}
\caption{Robustness of the gate fidelity $\mathcal{F}$ for dCRAB-optimised two-photon ARP $CCZ$ pulses of \fref{gates_optimal} (a) $1-\mathcal{F}$ for a range of relative variations of $\pm 3\%$ on the laser intensity and $\pm 1\%$ on the two-photon detuning. (b) Rydberg interaction strength and fidelity as a function of the inter-atomic separation distance $d$ in an equilateral triangle configuration. (c) Average fidelity $\langle \mathcal{F}\rangle$ and standard deviation  $\sigma_\mathcal{F}$ obtained by running the gate protocol 150 times with Lorentzian laser linewidth with different values of the FWHM.}
\label{3Qgate_fidelities}
\end{figure}
After finding the optimal pulse sequence assuming ideal conditions, we now move on to evaluate the sensitivity of the $CCZ$ gate ARP protocol to different types of experimental imperfections. We start by considering fluctuations of the optimal parameters presented in the previous sections. In \fref{3Qgate_fidelities}(a) we show $1-\mathcal{F}$ obtained when the two-photon detuning $\delta$ and laser laser intensity $I_{1}^{e}$ are deviated from the optimal values used in \fref{gates_optimal} by a fixed relative amount throughout the pulse duration. For relative fluctuations of up to $\pm 3\%$ on the laser power and $\pm 1\%$ on the two-photon detuning, the fidelity remains in the range $0.9939 \leq \mathcal{F} \leq 0.9954$. In \fref{3Qgate_fidelities}(b) we plot $1-\mathcal{F}$ and the Rydberg interaction strength $V_{rr}$ as a function of the separation $d$ between the atoms in an equilateral triangle configuration.  The ARP pulses are the same as depicted in \fref{gates_optimal}, but for each interaction strength we have re-optimised the corrective phase $\phi$. The gate fidelity reduces slightly for increased separation due to finite Rydberg blockade, but remains $\mathcal{F}>0.993$ even for $d\sim 6~\mu$m. Independent evaluation of gate fidelity for geometries with asymmetric spacings between the atoms confirms that the performance of the protocol is limited by the minimum value amongst all the pairwise Rydberg interaction strengths rather than by the presence of unequal interaction strengths. 
\\
Furthermore, the performance of the gate is affected by issues that are common to all neutral atom experiments.  One of the most important sources of error is dephasing in the excitation to the Rydberg state caused by the fact that the excitation lasers have a finite linewidth. We have performed numerical calculations accounting for this effect by allowing for random phases $\phi_1(t)$ and $\phi_2(t)$ on the Rabi frequencies associated with the $\ket{1}\leftrightarrow \ket{e}$ and $\ket{e}\leftrightarrow \ket{r}$ lasers, i.e., we have modified them as
\begin{equation}
 \Omega_1^{ f_e,m_{f_e}}\rightarrow e^{i\phi_1(t)}\Omega_1^{ f_e,m_{f_e}}, \Omega_{f_e,m_{f_e}}^{r}\rightarrow e^{i\phi_2(t)}\Omega_{f_e,m_{f_e}}^{r}.
\end{equation}
According to Eq. \eqref{OmegaR}, the effective Rabi frequency is modified as $\Omega_R\rightarrow \Omega_R e^{i(\phi_1(t)+\phi_2(t))}$.  In order to speed up the calculations needed to obtain significant statistical samples, we have computed the gate fidelity under the presence of laser phase noise using the effective model described in Appendix \ref{reducedmodel_appendix}. The phases $\phi_1(t)$ and $ \phi_2(t)$ are stochastic quantities generated according to a laser phase noise spectral density $S_{\phi}(f)$, where $f$ denotes the Fourier frequency.  The phase noise spectral density is related to the frequency noise spectral density as $S_{\delta\nu}(f)=f^2S_{\phi}(f)$, where $\delta\nu=\nu-\nu_0$ and $\nu_0$ is the main laser frequency. The spectral function $S_{\delta\nu}(f)$ can be used to compute the auto-correlation function, which is the Fourier transform of the laser line shape $S_E(\delta\nu)$, where $E$ denotes the electric field amplitude. Assumimg that the lasers are cavity-locked,  we can approximate $S_E(\delta\nu)$ by a Lorentzian distribution.  Under this assumption, the frequency noise spectral function is $S_{\nu}(f)=constant=h_0/\pi$ (white noise), where $h_0$ is the Full Width at Half Maximum (FWHM) of $S_E(\delta\nu)$ \cite{di2010simple}. 

In~\fref{3Qgate_fidelities}(c) we plot the average value of the fidelity $\langle \mathcal{F}\rangle$ and the corresponding standard deviation $\sigma_\mathcal{F}$ obtained by running the optimal gate protocol 150 times using Lorentzian laser line shapes $S_E(\delta\nu)$ with different values of the FWHM linewidth. Although for a FWHM of 1~kHz the average fidelity drops considerably, for widths $\lesssim100$~Hz (which are routinely achieved experimentally e.g. \cite{legaie18}) it approaches the value of $\mathcal{F}=0.9954$ that we obtained without considering phase noise.  Besides reducing the laser bandwidth further through filtering using a high-finesse cavity \cite{levine2018high}, another possible approach to mitigate the effects of laser phase noise is to engineer a frequency noise spectral density $S_{\delta\nu}(f)$ which instead of having a flat profile is minimal for Fourier frequencies $f$ around the effective Rabi frequency $\Omega_R$, in such a way that the phase fluctuations have a very different frequency than the population transfer and therefore the excitation to the Rydberg manifold experiences little dephasing \cite{deLeseluc2018analysis}. 
\\
Besides laser phase noise, the proposed gate is highly sensitive to differences in laser intensity at each atom due to relative positioning and alignment with respect to the global beam and the delocalisation of the atomic wavefunction within the traps. These issues are common to gates relying on global driving, and they can be mitigated by utilising a spatial light modulator or holographic optical element to convert the beam profile to a top-hat rather than Gaussian distribution (as demonstrated e.g. in \cite{ebadi2021quantum}), to achieve uniform intensity across the atoms making the scheme robust to small misalignments and the delocalisation of the atoms. An alternative strategy is to correct for small differences in the field intensities by applying independent phase corrections $Z(\phi)$ to each atom which would allow correction for accumulated phase errors, but in practise this would be challenging to tune and optimise experimentally when targeting fidelities of the order of the ones reported in this paper. 
\section{Conclusion and outlook}
\begin{figure}[t!]
\includegraphics[width=17.2cm]{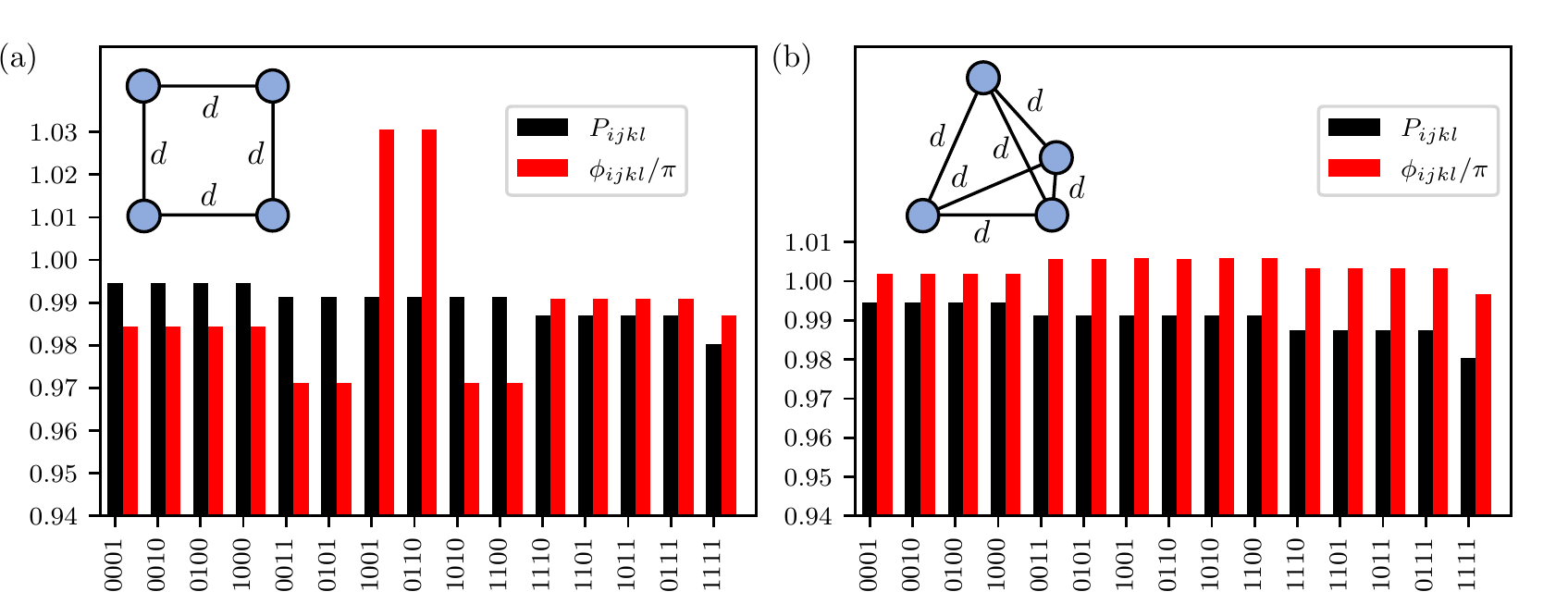}
\caption{Population transfer (black bars) and phase impriting (red bars) for all the input states coupled to the Rydberg manifold after the application of the $CCCZ$ gate protocol in (a) a square of side length $d=4$~$\mu$m and (b) a pyramid with all inter-atomic separations equal to 4~$\mu$m.  In both cases the applied pulses are the same as the analytical ones depicted in \fref{gates_normal}.}
\label{4Qgates}
\end{figure}

In conclusion, we have shown that ARP pulses can be implemented in a two-photon Rydberg excitation scheme with alkali atoms to realise a high fidelity $CCZ$ gate. Using a detailed model that takes into account all the relevant atomic levels and AC Stark shifts for Cs atoms, we have obtained $\mathcal{F}=0.9954$ with optimally shaped pulses found through dCRAB This gate is equivalent to 6 CNOT gates (plus additional single-qubit rotations) \cite{shende2008cnot}, meaning that matching the performance of our native $CCZ$ ARP gate would require the ability to perform two-qubit gates with a fidelity $\mathcal{F}\gtrsim (0.9954)^{1/6}=0.9992$ which so far remains elusive using two-photon excitation in alkali atoms (see Appendix \ref{decomposition_appendix}).
\\
The two-photon ARP controlled-phase protocol could in principle be generalised to any number of qubits by re-optimizing the pulse shapes. Using the analytical pulses of \fref{gates_normal} in a 4-qubit system formed by a square of size $d=4$~$\mu$m as shown in \fref{4Qgates}(a), we have found a gate fidelity $\mathcal{F}=0.9879$ using as a measure the generalisation of Eq. \eqref{fidelityformula} to a $CCCZ$ gate. In this case the relatively weak interaction $V_{rr}/2\pi=119$~MHz between atoms placed in opposite corners of the square leads to significant phase errors due to finite blockade.  In order to minimise this undesired effect, the ensemble could be arranged in a three-dimensional pyramidal configuration with three atoms in the $x-y$ plane and a fourth one equally separated from the triangle vertices, as shown in \fref{4Qgates}(b).  With all the inter-atomic distances set to $d=4~\mu$m in this geometry, the Rydberg interactions with the out-of-plane qubit are $V_{rr}/2\pi=612$~MHz, and the $CCCZ$ gate fidelity increases to $\mathcal{F}=0.9910$. Even though this fidelity would already be difficult to outperform with a decomposition of the $CCCZ$ gate into two- and three-qubit gates, further optimisation would be likely to lead to an improved performance. The main challenge in scaling the two-photon ARP gate scheme to an even larger number of qubits would be to ensure the condition that all the pairwise interactions are strong enough to provide total Rydberg blockade in the ensemble.

These results provide a viable route to scalable quantum computing on neutral atom arrays by enabling implementation of high-fideltiy multi-qubit unitiaries using parameters realistic for current experimental platforms. In future, higher fidelities may be attainable through power scaling to increase the coupling Rabi frequency from $\ket{e}\rightarrow\ket{r}$ to permit faster gate speeds and higher intermediate state detuning.

\begin{acknowledgments}
We thank Sebastian Weber for help with the integration of the ``pair interaction'' open-source software into our codes and Tomas Kozlej for providing numerical routines to evaluate laser phase noise. Furthermore, we are grateful to Callum Duncan and the members of the SQUARE project team for helpful discussions. This work is supported by the EPSRC (Grant No. EP/T005386/1) and M Squared Lasers Ltd. The data presented in this work are available at \cite{dataset}.
\end{acknowledgments}

%%%%%%%%%% Merge with supplemental materials %%%%%%%%%%
%\pagebreak
%\widetext
%\begin{center}
%\textbf{\large Supplementary information}
%\end{center}
%%%%%%%%%% Merge with supplemental materials %%%%%%%%%%
%%%%%%%%%% Prefix a "S" to all equations, figures, tables and reset the counter %%%%%%%%%%
%\setcounter{equation}{0}
%\setcounter{figure}{0}
%\setcounter{table}{0}
%\setcounter{page}{1}
%\makeatletter
%\renewcommand{\theequation}{S\arabic{equation}}
%\renewcommand{\thefigure}{S\arabic{figure}}
%\renewcommand{\citenumfont}[1]{S#1}
%%%%%%%%%% Prefix a "S" to all equations, figures, tables and reset the counter %%%%%%%%%%
\appendix

\section{Atomic physics parameters}
As stated in the main text, we consider the implementation of the two-photon ARP $C^kZ$ gate protocol in Cs atoms using the atomic states $\ket{0}=\ket{6S_{1/2},f=3,m_f=0}$, $\ket{1}=\ket{6S_{1/2},f=4,m_f=0}$, $\ket{e}=\ket{7P_{1/2}}$ and $\ket{r}=\ket{82S_{1/2},m_J=-1/2}$. The $\ket{1}\leftrightarrow\ket{e}$ transition is driven with a laser with $\sigma^+$ polarisation and detuned from $\ket{e}$ by $\Delta=16.3$ GHz, while the $\ket{e}$ and $\ket{r}$ states a coupled with a $\sigma^-$-polarised laser.  The Rabi frequencies of the drives are given by
\begin{equation}
\Omega_1^{f_e,m_{f_e}}=\sqrt{\frac{2I_1^e}{c\epsilon_0}}\frac{\langle 1\vert \hat{d} \vert f_e,m_{f_e} \rangle}{\hbar}, \Omega_{f_e,m_{f_e}}^r=\sqrt{\frac{2I_1^e}{c\epsilon_0}}\frac{\langle f_e,m_{f_e} \vert \hat{d} \vert r \rangle}{\hbar},
\end{equation}
where $I_1^e$ and $I_e^r$ are the intensities of the lasers driving the $\ket{1}\leftrightarrow\ket{e}$ and $\ket{e}\leftrightarrow\ket{r}$ transitions respectively, and $\hat{d}$ is the dipole operator. Details on how to compute efficiently the matrix elements between hyperfine and fine states using angular momentum algebra can be found in e.g.  \cite{maller2015rydberg}. Due to the dipole selection rules, the only hyperfine intermediate states for which $\Omega_1^{f_e,m_{f_e}}\neq 0$ are $\ket{7P_{1/2},f_e=3,m_{f_e}=1}$ and $\ket{7P_{1/2},f_e=4,m_{f_e}=1}$.  However, since we do not consider the hyperfine splitting of $\ket{r}$,  $\Omega_{f_e,m_{f_e}}^r\neq 0$ for all hyperfine components except $\ket{7P_{1/2},f_e=4,m_{f_e}=-4}$. We have used the Alkali Rydberg Calculator (ARC) open-source package \cite{sibalic17} to calculate all the relevant dipole matrix elements, or equivalently the ratios $\Omega_1^{f_e,m_{f_e}}/\sqrt{I_1^e}$ and $\Omega_{f_e,m_{f_e}}^r/\sqrt{I_e^r}$, as well as the hyperfine and Zeeman  splittings of the intermediate state, $E(f_e,m_{f_e})$. In table \ref{table:parameters} we report the values of $\Delta_1/I_1^e$,  $\Delta_r/I_e^r$ and $\Omega_R/\sqrt{I_1^eI_e^r}$ that we have computed with this procedure and used in all our calculations. We also provide the linewidths $\Gamma_e$ and $\Gamma_r$ of the states $\ket{e}$ and $\ket{r}$ at $T=0$ K, which we have also found with ARC and used throughout all our calculations.  For the computation of the dipole Rydberg interactions we have used the open source package pairinteraction \cite{weber2017calculation}. All the values of the interactions used for the calculations are reported in the main text.
\begin{table}[h!]
\centering % used for centering table
\begin{tabular}{ c  c } % centered columns (2 columns)
\hline \hline%inserts horizontal line
Parameter & Value  \\ [0.5ex] % inserts table
\hline
$\Delta_1/I_1^e$ &$2\pi\times 24.006$~MHz~$\mu$m$^2$/mW \\ [2ex]% inserting body of the table
$\Delta_r/I_e^r$ &$2\pi\times 0.643$~MHz~$\mu$m$^2$/mW\\ [2ex]% inserting body of the table
$\Omega_R/\sqrt{I_1^eI_e^r}$ & $2\pi\times 2.780$~MHz~$\mu$m$^2$/mW\\ [2ex]
$\Gamma_e$ &$2\pi\times 1.031$~MHz \\ [2ex]
$\Gamma_r$ &$2\pi\times 0.280$ kHz \\ [1ex] % [1ex] adds vertical space
\hline \hline
\end{tabular}
\caption{Summary of the relevant atomic physics parameters used in our calculations. The data corresponds to an intermediate-state detuning $\Delta=16.3$ GHz and a magnetic field $B=10$ G perpendicular to the atomic plane.} % title of Table
\label{table:parameters} % is used to refer this table in the text
\end{table}

\section{Reduced model without intermediate states}
\label{reducedmodel_appendix}
\begin{figure}[h!]
\includegraphics[width=11.7cm]{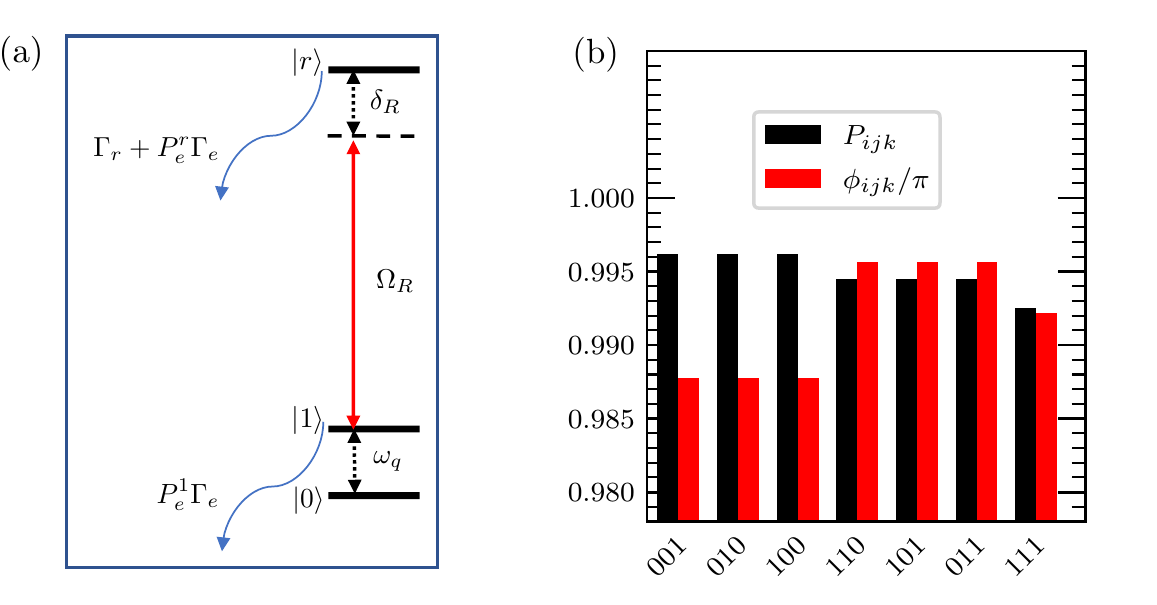}
\caption{(a) Sketch of the atomic levels and parameters considered in the reduced model described by the single-atom Hamiltonian \eqref{Ham_reduced}. (b) Population transfer (black bars) and phase imprinting (red bars) obtained with the reduced model without intermediate states for the dCRAB-optimised pulses presented in the main text.}
\label{reducedmodel}
\end{figure}
In order to reduce the computational cost of optimizing the pulse shapes, we employ for this purpose a reduced model in which the intermediate excited states are eliminated while retaining an accurate description of their dissipative effect.  Using the adiabatic elimination procedure, we can estimate the total population of the intermediate states as
\begin{equation}
P_e(t)=\sum_{f_e,m_{f_e}}\frac{\vert\Omega_1^{ f_e,m_{f_e}}(t)\vert^2+\vert\Omega_r^{ f_e,m_{f_e}}(t)\vert^2}{4(\Delta_{f_e,m_{f_e}})^2}=P_e^1(t)+P_e^r(t),
\end{equation}
where $P_e^1(t)=\sum_{f_e,m_{f_e}}\vert\Omega_1^{ f_e,m_{f_e}}(t)\vert^2/4(\Delta_{f_e,m_{f_e}})^2$ and $P_e^r(t)=\sum_{f_e,m_{f_e}}\vert\Omega_r^{ f_e,m_{f_e}}(t)\vert^2/4(\Delta_{f_e,m_{f_e}})^2$ are the contributions to the intermediate-state population coming from the laser couplings from $\ket{1}$ to $\ket{e}$ and from $\ket{e}$ to $\ket{r}$, respectively.  Thus, we can incorporate the effects of spontaneous decay from $\ket{e}$ into the effective single-atom Hamiltonian describing the coupling from $\ket{1}$ to $\ket{r}$ by introducing effective dissipation terms from $\ket{1}$ and $\ket{r}$ weighed by $P_e^1$ and $P_e^r$, respectively. This effective Hamiltonian, the parameters of which are depicted in \fref{reducedmodel}(a) (note that $\delta_R=\delta+\Delta_1-\Delta_r$), reads
\begin{equation}
\frac{\mathcal{H}_{\text{c/t}}^\text{eff}}{\hbar}=\left(\Delta_1-i\frac{P_e^1\Gamma_e}{2} \right)\ket{1}\bra{1}+\frac{\Omega_R}{2}\left(\ket{1}\bra{r}+\ket{r}\bra{1}\right)+\left[(\Delta_r-\delta)-\frac{i}{2}\left(\Gamma_r+P_e^r\Gamma_e\right)\right]\ket{r}\bra{r}.
\label{Ham_reduced}
\end{equation} 
This effective model yields almost identical results to the full model explicitly taking into account all the intermediate excited states. In \fref{reducedmodel}(b) we show the population transfer and phase imprinting computed with the effective model for the dCRAB-optimised pulses discussed in the main text and illustrated in \fref{gates_optimal} of the main text.  For this specific pulse sequence, the relative difference in the gate fidelities computed with the exact and effective model is $\vert(\mathcal{F}-\mathcal{F}_{\text{eff}})/\mathcal{F}\vert=4\times 10^{-6}$. This approach thus provides a robust method to accurately model gate fidelities in a reduced Hilbert space to enable efficient optimisation and permit scaling to larger numbers of qubits.

\section{Optimal pulse parameters} \label{optimisation}
We have applied {a variant of the dCRAB quantum optimisation technique to find an improved shape for the effective Rabi frequency $\tilde{\Omega}_R=g(t)\Omega_R(t)$, where $\Omega_R(t)$ is the optimised analytical pulse described in the main text and depicted in \fref{gates_normal}. The envelope function $g(t)$ has the form
\begin{equation}
g(t)= 
\begin{cases} 
     1+\sum_{k=1}^{N}A_k \cos \left(\omega^1_kt\right) +B_k \sin \left(\omega^1_kt\right);0\leq t \leq T_1 \\
      \\
     1+\sum_{k=1}^{N}A_k \cos \left(\omega^2_k (t-T_1)\right) +B_k \sin \left(\omega^2_k (t-T_1)\right);T_1< t \leq T_2\\
   \end{cases},
\label{envelopeCRAB}
\end{equation}
with $\omega^i_k=2\pi k r_k/T_i$.  We have set a frequency cut-off at $N=6$ and performed the numerical optimisation with respect to the variables $A_k, B_k$ and $r_k$.  The optimal parameters that we have found with this method and then used in the calculations of Figs. 3 and 4 of the main text are summarised in Table \ref{table:parameters_CRAB}.
\begin{table}[h!]
\begin{tabular}{c c c c c c c} % centered columns (2 columns)
\hline \hline%inserts horizontal line
Parameter & $k=1$ & $k=2$ & $k=3$ & $k=4$ & $k=5$ & $k=6$   \\ [0.5ex] % inserts table
\hline
$A_k$ & -0.0859 & 0.0145 & 0.3612 & -0.2605 & 0.4847 & 0.05360\\ [2ex]
$B_k$ & -0.7250 & -1.7963 & 0.9775 & -0.4293 & 0.5623 & -0.5406\\ [2ex]
$r_k$ & 0.3930 & 0.0402 & 0.0597 & 0.3959 & -0.2616 & -0.2132\\ [1ex] % [1ex] adds vertical space
\hline \hline
\end{tabular}
\caption{Numerical values of the optimal parameters of the dCRAB envelope function given by eq. \eqref{envelopeCRAB}.} 
\label{table:parameters_CRAB} % is used to refer this table in the text
\end{table}

\section{Decomposition of $CCZ$ gate into $CZ$ and single-qubit gates}
\label{decomposition_appendix}
In \fref{circuitdecomp} we depict the circuit representation of the optimal decomposition of the $CCZ$ gate into six $CZ$ gates and single-qubit gates \cite{shende2008cnot}. The single-qubit rotation is given $T=\exp(i\pi/8\sigma_z)$, and $H$ represents the Hadamard gate. In order for this decomposition to work with the definition of the $C^kZ$ gates that we have employed in this paper $U_{C^{k}Z}=2\left(\otimes_{k+1}\ket{0} \otimes_{k+1}\bra{0}\right)-\mathds{I}$, the Hadamard gate needs to be defined with the roles of the $\ket{0}$ and $\ket{1}$ states swapped, $H=1/\sqrt{2}(-\ket{0}\bra{0}+\ket{0}\bra{1}+\ket{1}\bra{0}+\ket{1}\bra{1})$.  Note that the $CCZ$ could in principle be implemented with five two-qubit gates \cite{barenco95}, but this would require realizing a $CV=(1-i)\ket{00}\bra{00}-\mathds{I}$ gate which is not native to Rydberg systems.
\begin{figure}[h!]
\includegraphics[width=13.8cm]{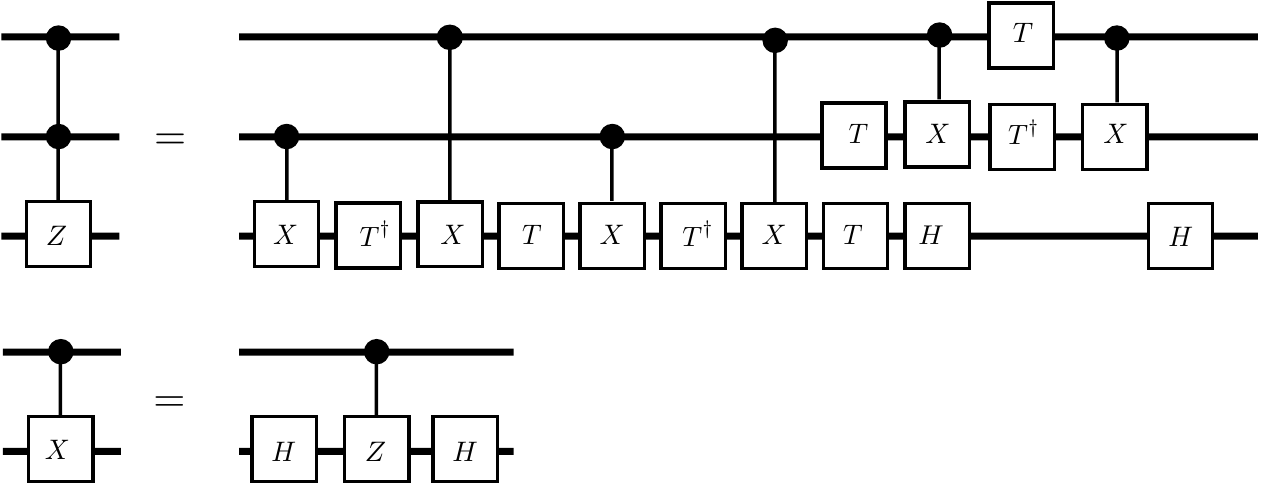}
\caption{Quantum circuit representation of a decomposition of the $CCZ$ into the minimum possible number of $CZ$ gates and additional single-qubit gates. Adapted from \cite{shende2008cnot}.}
\label{circuitdecomp}
\end{figure}
Using the ARP protocol outlined in the paper we optimise the two-qubit $CZ$ gate pulse sequence for equivalent parameters to those used for the $CCZ$ gate. The two-qubit gate fidelity is calculated as above using
\begin{equation}
\mathcal{F}_{2Q}=\left\vert \left\langle\Psi\vert U_{CZ}^{\dagger}U \vert \Psi\right\rangle \right\vert^2,
\end{equation}
where $U_{CZ}^{\dagger}$ is the ideal $CZ$ matrix, $U$ is the non-unitary operator implemented by the actual pulses and $\vert\Psi\rangle$ is a symmetric superposition of the computational input states. We obtain an optimal two-qubit gate fidelity $\mathcal{F}_{2Q}\sim 0.9985$ for a total pulse time of $\tau\sim 1.3$~$\mu$s. We compare this performance to the sequential $CZ$ pulse sequence presented in \cite{jaksch2000fast,maller2015rydberg} which requires time $\tau=4\pi/\Omega_R\sim 0.5~\mu$s and the global addressing $CZ$ gate scheme of \cite{Levine2019parallel} which requires $\tau\sim 2.73\pi/\Omega_R\sim0.3~\mu$s, and obtain comparable performance with $\mathcal{F}_{2Q}\sim 0.9985$ for all three gate schemes, which exceed the performance of two-photon protocols based on STIRAP \cite{saffman2020symmetric}.

Using these optimal two-qubit $CZ$ gates we simulate the $CCZ$ gate decomposition circuit depicted in \fref{circuitdecomp} assuming perfect single-qubit gates, and find $CCZ$ gate fidelities $\mathcal{F}\sim 0.991$, which are significantly lower than the fidelity $\mathcal{F}= 0.9954$ obtained with the optimal two-photon ARP pulses discussed in the main text. In order for the two-qubit gates to outperform the our native $CCZ$ ARP gate, their fidelity would need to be $\mathcal{F}_{2Q}\sim(0.9954)^{1/6}\sim 0.9992$. This fidelity exceeds any known protocol based on two-photon excitation, and are only predicted for direct single-photon transitions using UV wavelengths \cite{saffman2020symmetric,theis2016high} which introduces technical challenges due to power scaling and increased Doppler sensitivity. However, even if these excellent fidelities were achievable, the native $CCZ$ would still have the advantage of requiring only a single-qubit rotation and to have a total duration shorter than the quantum pulse sequence associated to the quantum circuit depicted in \fref{circuitdecomp}. 
%\section{4. Four-qubit gate}
%Provide more details about the non-optimised four-qubit gate calculation mentioned in the main text.   
\bibliography{multiqubitARP_QST_revised.bbl}
\end{document}